\documentclass[aps,pra,nofootinbib]{revtex4-2}

\usepackage{graphicx,amsmath,amssymb}

\usepackage{dcolumn}

\usepackage{color,bm}

\newcommand{\tcr}[1]{\textcolor{black}{#1}}

\usepackage{bm}

\begin{document}

\title{Heralding probability optimization for nonclassical light generated by photon counting measurements on multimode Gaussian states}

\author{Jaromír Fiurášek}

\email[Corresponding author: ]{fiurasek@optics.upol.cz}
\affiliation{Department of Optics, Faculty of Science, Palack\'y University, 17.\ listopadu 12, 77900  Olomouc, Czech Republic}

\begin{abstract} 
Generation of highly nonclassical quantum states of light is essential for optical quantum information processing and quantum metrology. Given the lack of sufficiently strong nonlinear interactions between optical fields, the commonly employed optical quantum-state preparation schemes are conditional, based on nonlinearity induced by heralding photon number measurement on a part of a multimode squeezed Gaussian state. The development and optimization of such probabilistic quantum-state engineering schemes represents one of  the central challenges in current quantum optics. As technology advances and experiments progress toward the detection of higher numbers of photons, the maximization of the heralding probability becomes essential to ensure sufficiently high state-preparation rates. Here, we show that for  conditional quantum state preparation schemes based on Gaussian states and photon number measurements, the maximization of the heralding probability can be achieved by finding solutions to a system of polynomial equations, which offers an efficient way to find the optimal configuration and allows us to apply techniques dedicated specifically to solving such systems of equations. Our approach can seamlessly incorporate bounds on available single-mode quadrature squeezing, which is highly experimentally relevant. We mainly consider the generation of finite superpositions of Fock states but show that the approach can be straightforwardly extended to the generation of squeezed superpositions of Fock states. We focus on Gaussian states with vanishing coherent displacements, hence, the conditionally generated states have well-defined photon number parity. We illustrate our general methodology through examples of generation of single-mode states with two and three heralding modes, and generation of two-mode states with two heralding modes.

\end{abstract}

\maketitle

\section{Introduction}

Generating highly nonclassical, \tcr{non-Gaussian} quantum states of light, such as Fock states  \cite{Lvovsky2001,Cooper2013,Bouillard2019,Kawasaki2022}, Gottesman-Kitaev-Preskill states \cite{Takase2023,Konno2024,Larsen2025}, superpositions of coherent states \cite{Ourjoumtsev2006,Nielsen2006,Ourjoumtsev2007,Wakui2007,Takahashi2008,Yukawa2013,Huang2015,Sychev2017,Etesse2015,Cotte2022,Simon2024,Endo2025,Yu2026}, or states exhibiting nonlinear squeezing  \cite{Konno2021,Kala2025} is essential for optical quantum computing, quantum error correction, and quantum metrology and sensing \cite{Gottesman2001,Ralph2003,Giovanetti2011,Pirandola2018,Guillaud2019,Brady2024,Tzitrin2021,Albert2018,Schegel2022,Grochowski2025}. Arguably the most popular and fruitful approach to engineering such sophisticated non-Gaussian quantum states of optical fields \cite{Walschaers2021} consists of conditional heralding schemes, where photon counting measurements on a part of multimode Gaussian state herald the successful preparation of a non-Gaussian state in the remaining modes \cite{Lvovsky2020}. Such conditional quantum state generation schemes are closely related to Gaussian boson sampling \cite{Lund2014,Hamilton2017,Paesani2019,Zhong2020,Zhong2021,Thekkadath2022} and they encompass a large number of both early and recent experiments in the field, including  conditional photon addition and (generalized) photon subtraction \cite{Zavatta2004,Ourjoumtsev2006,Nielsen2006,Wakui2007,Nielsen2010,Fadrny2024,Tomoda2024,Takase2024,Chen2024}.

Recently, the optimization of such conditional quantum-state generation schemes with respect to the properties of the generated states, as well as the maximization of the heralding probability, has gained significant attention \cite{Su2019,Mogyorosi2019,Hanamura2025,Fiurasek2025arXiv,Notarnicola2026}. \tcr{The heralding probability is the probability that the conditional state preparation succeeds; that is, the probability that measurements performed on the heralding modes yield the required outcomes.} Importantly, the optimization of the structure of the generated quantum state and the maximization of the heralding probability can be largely separated. The parameters that characterize the input Gaussian state can be divided into two groups, and the so-called damping parameters influence only the heralding probability but not the shape of the generated state \cite{Hanamura2025,Fiurasek2025arXiv}. Maximization of the heralding probability is crucial as the experiments advance to higher numbers of heralding photons and generation of states with complex structure in phase space \cite{Larsen2025,Yu2026,Takase2026}.

Although generic optimization algorithms can be utilized to maximize the heralding probability, it turns out that this problem can be formulated as finding solution to a system of polynomial equations, \tcr{which is a broadly studied problem in mathematics and numerical computation. Dedicated techniques and approaches that have been developed to solve systems of polynomial equations,}  such as the Gröbner basis construction or homotopy continuation \tcr{\cite{Li1997,Sturmfels2002,Bates2023}}, can be applied  to efficiently find  the optimal configuration that maximizes the heralding probability. Here, we develop this approach in detail for input pure Gaussian states with vanishing coherent displacements. For this class of input Gaussian states, the photon counting measurements generate states with well defined photon number parity, which encompasses important classes of states such as GKP states or cat states. Our calculations benefit greatly from the stellar (or Bargmann) representation of the quantum states of bosonic systems and the concept of core states \cite{Motamedi2025}. 

 Our approach can seamlessly incorporate a bound on the available quadrature squeezing which is highly practically relevant, because the unconstrained optimization may yield squeezing levels that are not yet experimentally available. We mostly assume that the determination of parameters that specify the generated state and the maximization of the heralding probability can be separated, but we show that our approach is general enough to handle even the case when these to tasks become coupled and intertwined. The main contribution of the present work is methodological, nevertheless, we illustrate the general methodology on explicit examples with two \tcr{and three} heralding modes.  

The \tcr{rest} of the paper is organized as follows. In Section II, we present the general formalism and derive extremal equations for the damping parameters that maximize the heralding probability. We also demonstrate how to handle limited available squeezing. In Section III, we extend the formalism to the generation of squeezed superpositions of Fock states. In Section IV, we apply the formalism to the generation of single-mode non-Gaussian states by photon counting measurements on two heralding modes. \tcr{In Section V, we provide examples of optimization of the herealding probability for a scheme with three heralding modes.}  Next, in Section VI, we give a simple example of the maximization of the heralding probability for target two-mode non-Gaussian state, a single-rail entangled two-qubit state. This example not only shows that the present formalism is applicable to the generation of multimode states, but also illustrates the  situation where some extra parameters besides the damping factors become free and available for optimization.  Finally, Section VII contains a brief summary and conclusions.

\begin{figure}[!t!]
\includegraphics[width=0.43\linewidth]{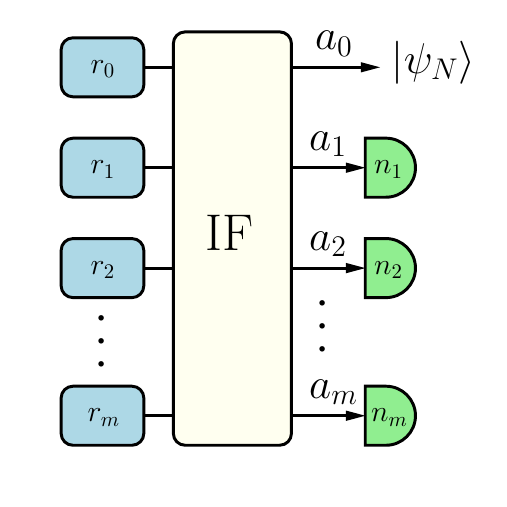}
\caption{Scheme for the generation of a single-mode nonclassical state of light $|\psi_N\rangle$ with stellar rank $N$ by photon counting measurements on $m$ heralding modes of $m+1$-mode Gaussian state. The Gaussian state has vanishing coherent displacements and can be generated from $m+1$ single-mode squeezed vacuum states with suitably chosen squeezing constants $r_j$ by their interference in an optical interferometer IF.}
\label{figscheme}
\end{figure}

\section{Maximization of the heralding probability - general procedure}
The considered quantum state generation scheme is shown in Fig.~\ref{figscheme}. An $m+1$-mode pure Gaussian quantum state $|G\rangle$ is generated by interference of $m+1$ single-mode squeezed vacuum states with squeezing constants $r_j$ in an optical interferometer IF \cite{Braunstein2005} and the $m$ heralding modes are measured in Fock basis. We associate the creation operator $\hat{a}_0$ with the signal mode where the output state is generated, and $\hat{a}_j^\dagger$, $1 \leq  j\leq m$, denote the creation operators of the heralding modes. Furthermore, we denote by $n_j$ the detected number of photons in the $j$th heralding mode.

\tcr{The heralding probability (also referred to as the success probability) $p_S$ is defined as the probability of successful detection of the specific chosen  heralding photon number pattern $(n_1, n_2, \ldots, n_m)$, 
\begin{equation}
p_S= \mathrm{Tr}\left[\left( \hat{I}_{0}\otimes |\bm{n}\rangle\langle \bm{n}|\right)| G\rangle \langle G|\right].
\end{equation}
Here $\hat{I}_0$ denotes the identity operator on the signal mode $a_0$ and $|\bm{n}\rangle=|n_1,n_2,\ldots,n_m\rangle$ is a short-hand notation for  multimode  Fock state of the $m$ heralding modes. The non-normalized conditionally generated state in the signal mode $a_0$ can be expressed as
\begin{equation}
|\tilde{\psi}_N\rangle= \langle \bm{n}|G\rangle,
\end{equation}
and the heralding probability can be calculated as $p_S=\langle \tilde{\psi}_N|\tilde{\psi}_N\rangle$.} \tcr{Here $N=\sum_{j=1}^m n_j$ denotes the total number of the detected photons in the heralding modes.}

\tcr{In order to derive explicit expression for the heralding probability $p_S$ we make use of the stellar (or Bargmann) representation of quantum states.}
Any pure $m+1$-mode Gaussian state $|G\rangle$  with vanishing coherent displacements can be expressed as \cite{Motamedi2025,Fiurasek2025arXiv}
\begin{equation}
|G\rangle =Z^{1/4} \exp\left(\frac{1}{2} \sum_{j=0}^m \sum_{k=0}^m A_{jk} \hat{a}_j^{\dagger}\hat{a}_k^\dagger\right) |\mathrm{vac}\rangle,
\label{Gstatedefinition}
\end{equation}
where $A$ is a symmetric complex matrix, and $|\mathrm{vac}\rangle$ denotes the vacuum state of all $m+1$ modes. The normalization factor $Z$ is given by \cite{Fiurasek2025arXiv}
\begin{equation}
Z=\det(I-AA^\dagger),
\label{Zdefinition}
\end{equation}
and the matrix $A$ must satisfy the inequality 
\begin{equation}
I-AA^\dagger >0,
\label{physicalitycondition}
\end{equation}
 where $I$  denotes the  identity matrix.  The physicality condition
 (\ref{physicalitycondition})  guarantees that the Gaussian  state $|G\rangle$ is normalizable, possesses a well-behaved covariance matrix  and exhibits finite quadrature squeezing. 
 The singular values of $A$ read $\tanh r_j$ hence they specify the single-mode squeezings required to generate the state $|G\rangle$. The expression (\ref{Gstatedefinition}) is closely related to the stellar representation of the state $|G\rangle$ \cite{Vourdas2006,Motamedi2025}, which is defined via overlap of the state $|G\rangle$ with multimode coherent state $|z_0^\ast,z_1^\ast,\ldots,z_m^\ast\rangle$ with complex amplitudes $z_j^\ast$,
 \begin{equation}
 f(\bm{z})= \prod_{j=0}^m e^{\frac{|z_j|^2}{2}} \langle z_0^\ast,z_1^\ast,\ldots,z_m^\ast|G\rangle.
 \end{equation}
For the Gaussian state (\ref{Gstatedefinition}) we obtain 
\begin{equation}
f_G(\bm{z})=Z^{1/4} \exp\left(\frac{1}{2} \sum_{j=0}^m \sum_{k=0}^m A_{jk} z_j z_k\right).
\label{fzdefinition}
\end{equation}
\tcr{ Fock state $|n\rangle$ can be formally obtained from a coherent state $|\alpha\rangle$ as follows,
\begin{equation}
|n\rangle= \left.\frac{1}{\sqrt{n!}} \frac{\partial ^n}{\partial \alpha^n} \left[ e^{\frac{|\alpha|^2}{2}}|\alpha\rangle\right] \right|_{\alpha=0}.
\end{equation}
Therefore, the stellar function $g_\psi(z_0)$ of the non-normalized conditionally generated non-Gaussian state $|\tilde{\psi}_N\rangle$ can be expressed as
\begin{equation}
g_\psi(z_0)= \left.\prod_{j=1}^m \frac{1}{\sqrt{n_j!}}\frac{\partial^{n_j}}{\partial z_j^{n_j}} f_G(\bm{z}) \right|_{z_1=z_2=\ldots=z_m=0}. 
\end{equation}
}

In what follows, we  specifically consider the so-called  Gaussian core states $|G\rangle$ that satisfy $A_{00}=0$. Furthermore, we assume that the signal mode is directly coupled to all heralding modes and $A_{0j} \neq 0$, $j>0.$ For such a core state, the conditionally generated state in mode $a_0$ is a finite superposition of Fock states up to  $N=\sum_{j=1}^m n_j$ \cite{Motamedi2025},
\begin{equation}
|\psi_N\rangle=\sum_{k=0}^N c_k |k\rangle.
\label{psidefinition}
\end{equation}
The stellar rank \cite{Chabaud2020,Chabaud2021,Lachman2019} of the generated non-Gaussian state is thus determined by the  total number of photons counted in the heralding modes and is equal to $N$. The displacement-free Gaussian state (\ref{Gstatedefinition}) has even parity with respect to the total photon number. This implies that the conditionally generated state $|\psi_N\rangle$ will exhibit either even or odd parity in Fock  basis, determined by the parity of $N$. Consequently, for even $N$, we have $c_{2k+1}=0$, whereas for odd $N$ we have $c_{2k}=0$. 

\tcr{ Since the conditionally generated state (\ref{psidefinition}) is a finite suprposition of Fock states, the stellar function $g_{\psi}(z_0)$ is a polynomial of degree $N$.
To see the connection between the Fock-state amplitudes $c_k$ and the stellar polynomial $g_\psi(z_0)$, it is convenient to represent the state (\ref{psidefinition}) as a polynomial in creation operator $\hat{a}_0^\dagger$ acting on the vacuum,}
\begin{equation}
|\psi_N\rangle \propto \sum_{k=0}^N d_k \hat{a}_{0}^{\dagger k} |0\rangle, 
\label{psiNdkexpansion}
\end{equation}
where 
\begin{equation}
d_k= \sqrt{\frac{N!}{k!}} \frac{c_k}{c_N}.
\label{dkck}
\end{equation}
The coefficients  $d_k$ ($c_k$) depend only on a specific subset of parameters that specify the matrix $A$, because any mathematical  scaling $\hat{a}_j^\dagger\rightarrow x_j\hat{a}_j^\dagger$, $j>0$,
changes only the heralding probability but not the form of the conditionally generated state \cite{Hanamura2025}. Therefore, we can decompose the matrix $A$ as follows,
\begin{equation}
A= \Lambda B \Lambda,
\end{equation}
where $\Lambda_{jk}=x_j \delta_{jk}$  is a real positive-definite diagonal matrix with $x_0=1$, and the elements of the complex symmetric matrix $B$ can be parameterized as follows,
\begin{equation}
B_{00}=0, \quad B_{0j}=B_{j0}=1, \quad B_{jj}=s_j, \quad B_{jk}=B_{kj}=\nu_{jk}, \quad 1 \leq  j,k\leq m, \quad j\neq k .
\label{Belements}
\end{equation}
Explicitly, the matrices  $B$ and $\Lambda$ read,
\begin{equation}
\Lambda=\left(
\begin{array}{cccc}
1 & 0 & \cdots & 0 \\
0 & x_1 & \ddots & \vdots \\
\vdots & \ddots & \ddots & 0 \\
0 & \cdots & 0 & x_m
\end{array}
\right),
\qquad
B=\left(
\begin{array}{ccccc}
0 & 1 & 1 & \cdots & 1 \\
1 & s_1 & \nu_{12} & \cdots & \nu_{1m} \\
1 &\nu_{12} & s_2 & \cdots & \vdots \\
\vdots & \vdots & \ddots & \ddots & \nu_{m-1,m} \\
1 & \nu_{1m} & \cdots & \nu_{m-1,m} & s_m 
\end{array}
\right),
\label{BLambdaexplicit}
\end{equation}
\tcr{Making use of the decomposition $A=\Lambda B\Lambda$  and the Gaussian form of the stellar function $f_G(\bm{z})$, we get
\begin{equation}
g_\psi(z_0)=\left. \left[Z^{1/4} \prod_{k=1}^m \frac{x_k^{n_k}}{\sqrt{n_k!}}\right] \prod_{j=1}^m \frac{\partial^{n_j}}{\partial z_j^{n_j}} \exp \left(\frac{1}{2}\sum_{u=0}^m\sum_{v=0}^m B_{uv}z_u z_v\right)\right|_{z_1=\ldots z_m=0}.
\end{equation}
Performing the differentiation yields a polynomial function of $z_0$,
\begin{equation}
g_\psi(z_0)= Z^{1/4} \prod_{j=1}^m \frac{x_j^{n_j}}{\sqrt{n_j!}} \sum_{k=0}^N {d_k} z_0^k.
\label{gpsipolynom}
\end{equation}}
\tcr{Consequently, the non-normalized conditionally generated state $|\tilde{\psi}_N\rangle$ can be written in Fock basis as 
\begin{equation}
|\tilde{\psi}_N\rangle=Z^{1/4} \prod_{j=1}^m \frac{x_j^{n_j}}{\sqrt{n_j!}} \sum_{k=0}^N {d_k} \sqrt{k!}|k\rangle.
\label{psiexplicit}
\end{equation}}
The coefficients $d_k$ that specify the conditionally generated state  depend only on the elements of matrix $B$  and can be expressed as 
\begin{equation}
d_k=\left.\frac{1}{k!} \frac{\partial ^k}{\partial z_{0}^k} \prod_{j=1}^m \frac{\partial^{n_j}}{\partial z_j^{n_j}} \exp \left(\frac{1}{2}\sum_{u=0}^m\sum_{v=0}^m B_{uv}z_u z_v\right) \right|_{z_0=z_1=\ldots z_m=0}.
\label{dkdefinition}
\end{equation}
Observe that  $d_k$ are polynomial functions of matrix elements of $B$. \tcr{We can therefore call the parameters $s_j$ and $\nu_{jk}$ the control parameters, following the terminology of Ref. \cite{Hanamura2025}.}  More specifically, each nonzero $d_k$ is a homogeneous polynomial of degree $(N-k)/2$ \tcr{in the control parameters}, and $d_N=1$ by construction.

\tcr{In order to write down the final formula for the heraliding probability $p_S$, we make use of the identity
\begin{equation}
\frac{N!}{|c_N|^2}=\sum_{k=0}^N k! |d_k|^2.
\end{equation}
We combine this identity with the formula (\ref{psiexplicit}) and the expression $p_S=\langle \tilde{\psi}_N|\tilde{\psi}_N\rangle$ to obtain}
\begin{equation}
p_S=\frac{N!}{|c_N|^2} \left[\det(I-AA^\dagger)\right]^{1/2}\prod_{j=1}^m \frac{x_j^{2n_j}}{n_j!}.
\label{pSdefinition}
\end{equation}
The heralding probability $p_S$ can be maximized by optimizing the free parameters that are not fixed by the choice of the target state $|\psi_N\rangle$. Let us first assume that all the matrix elements of $B$ are fixed by the choice of $|\psi_N\rangle$, and we optimize the parameters $x_j$. As a preparatory step, we switch to new variables $X_j=x_j^2$. The determinant in equation (\ref{pSdefinition}) is a polynomial function of $X_j$. This follows from the identity
\begin{equation}
\det(I-AA^\dagger)=\det \left(I-\Lambda B \Lambda^2 B^\dagger \Lambda\right)=
\det \left(\Lambda\Lambda^{-1}-\Lambda B \Lambda^2 B^\dagger \Lambda^2\Lambda^{-1}\right)= 
 \det \left(I-B\Lambda^2 B^\dagger \Lambda^2\right ).
 \label{detidentity}
\end{equation}
These equalities hold because $\Lambda$ is invertible, the determinant of the  product of matrices is equal to the product of determinants, and $\det \Lambda \det\Lambda^{-1}=1$.  Equation (\ref{detidentity}) motivates the introduction of a new diagonal matrix $K=\Lambda^2=\mathrm{diag}(1,X_1,\ldots,X_m)$. The maximization of $p_S$ is equivalent to maximization of 
\begin{equation}
\tilde{p}=\det\left(I-B K B^\dagger K\right ) \prod_{j=1}^m X_j^{2n_j},
\label{tildepdefinition}
\end{equation}
where we have avoided the square root by considering $p_S^2$ and  omitted the terms that do not depend on $X_j$. The extremal equations for the optimal $X_j$ read
\begin{equation}
\frac{\partial\tilde{p}}{\partial X_j}=0.
\label{Xjextremal}
\end{equation}
We assume that $n_j>0$. Therefore, the optimal $X_j$ must satisfy $X_j>0$. Making use of the Jacobi's formula for derivative of a matrix determinant,
\begin{equation}
 \frac{d}{dt}\det M=(\det M) \mathrm{Tr}\left(M^{-1} \frac{d M}{dt} \right), 
 \label{Jacobi}
 \end{equation}
 we arrive at the following algebraic equations for the optimal $X_j$,
\begin{equation}
Q_j= \det(I-BKB^\dagger K)\left\{2n_j-X_j\mathrm{Tr}\left[(I-BKB^\dagger K)^{-1}(B\Sigma_jB^\dagger K+BKB^\dagger \Sigma_j)\right]\right\}=0.
\label{Qextremal}
\end{equation}
Here $\Sigma_j$ is a matrix whose  elements are all equal to zero except for one diagonal element, $(\Sigma_j)_{kl}=\delta _{kj}\delta_{lj}$ . The extremal equations (\ref{Qextremal}) form a system of $m$ polynomial equations for $m$ variables $X_j$. Such a system can be converted into a polynomial equation for a single variable, e.g., via the construction of a Gröbner basis \cite{Sturmfels2002}. However, the degree and complexity of the resulting polynomial equation may increase rapidly with the size of the original problem. For larger systems, the homotopy continuation method can be applied to determine the roots numerically \cite{Li1997,Bates2023}.  \tcr{Importantly, these methods are capable to identify all the roots of the system of polynomial equations (\ref{Qextremal}).  Therefore, global optimization of the heralding probability over all physically admissible values of $X_j$ is possible by checking all the physical solutions that satisfy $X_j>0$ and  the inequality (\ref{physicalitycondition}), and choosing the one that maximizes $p_S$.}

If $B$ is real, then $B^\dagger=B$ and the determinant in equation (\ref{tildepdefinition}) factorizes,
\begin{equation}
\det(I-BKBK)=\det(I-BK) \det(I+BK).
\end{equation}
It follows from the structure of the extremal equations (\ref{Xjextremal}) that all these equations will be satisfied when 
\begin{equation}
\det(I-BK)=\det(I+BK)=0.
\label{detsextremal}
\end{equation}
 For $m\geq 3$ this implies the existence of infinitely many solutions of the extremal equations (\ref{Qextremal}). However, all  solutions obtained from the conditions (\ref{detsextremal}) are unphysical, because they correspond to infinitely squeezed states. A quick numerical fix is to modify the function $\tilde{p}$ to 
\begin{equation}
\tilde{p}_\epsilon=\left[\det\left(I-B K B^\dagger K\right )+\epsilon \right] \prod_{j=1}^m X_j^{2n_j},
\label{pSepsilon}
\end{equation}
where $\epsilon$ is a sufficiently small parameter.  However, a more rigorous approach is possible, based on a suitable reformulation of the extremal equations (\ref{Xjextremal}) and (\ref{Qextremal}). Let us introduce a compact notation $E=\det(I-BK)$ and $F=\det(I+BK)$. When we take a suitable linear combination of the extremal equations (\ref{Xjextremal}) for $X_i$ and $X_j$, we obtain
\begin{equation}
\left( n_iX_j\frac{\partial F}{\partial X_j}-n_jX_i\frac{\partial F}{\partial X_i}\right)E +\left( n_i X_j\frac{\partial E}{\partial X_j}-n_jX_i\frac{\partial E}{\partial X_i}\right)F=0.
\label{difextremal}
\end{equation}
When equation (\ref{difextremal}) is considered for two different sets of indices $(i,j)$ and $(k,l)$, the resulting equations can be interpreted as a system of homogeneous equations for $E$ and $F$. Since we want solutions outside the unphysical boundary $E=0$ and/or $F=0$, we require that this system of homogenous equations has a nontrivial solution. This leads to the condition 
\begin{equation}
\tcr{W_{ij,kl}=}
\left|
\begin{array}{ccc}
\displaystyle{ n_iX_j\frac{\partial F}{\partial X_j}-n_jX_i\frac{\partial F}{\partial X_i}}  & ~~& \displaystyle{ n_i X_j\frac{\partial E}{\partial X_j}-n_jX_i\frac{\partial E}{\partial X_i} }\\[4mm]
\displaystyle{  n_kX_l\frac{\partial F}{\partial X_l}-n_lX_k\frac{\partial F}{\partial X_k}}  & ~~&   \displaystyle{n_k X_l\frac{\partial E}{\partial X_l}-n_lX_k\frac{\partial E}{\partial X_k}} 
\end{array}
\right|=0
\label{EFeliminated}
\end{equation}
where $||$ denotes the determinant of a $2\times 2$ matrix. We can combine the extremal equations (\ref{EFeliminated}) with the original extremal equations (\ref{Qextremal})  to obtain a system 
of polynomial equations \tcr{that may possess a} finite number of solutions when $B$ is real. 

In current experiments, the maximum available single-mode squeezing is limited. It is therefore highly relevant to maximize the probability of success under the constraint of limited available squeezing \cite{Larsen2025,Vahlbruch2016,Ha2026}. Let $r_{\mathrm{max}}$ denote the maximum achievable squeezing constant and define $\mu^2=\tanh^2 r_{\mathrm{max}}$. The squeezing constraint can be expressed as 
\begin{equation}
AA^\dagger \leq \mu^2 I.
\label{squeezingconstraint}
\end{equation}
When maximizing $p_S$ under this constraint, we must consider all the potential local maxima determined by solving the system of equations (\ref{Qextremal}), for which the constraint (\ref{squeezingconstraint}) is satisfied. In addition, we must maximize $p_S$ under the constraint that at least one of the eigenvalues of $AA^\dagger$ is equal to $\mu^2$, which results in the following condition:
\begin{equation}
D=\det\left(\mu^2 I- A A^\dagger\right)= \det\left(\mu^2 I-B K B^\dagger K\right)=0.
\label{Ddefinition}
\end{equation}
Note that this equality constraint takes the form of a polynomial equation for $X_j$. We can take this constraint into account by introducing the Lagrange multiplier $\lambda$ and optimizing the function  $\tilde{p}-\lambda D$. This yields a system of polynomial equations for $m+1$ variables $X_j$ and $\lambda$:
\begin{equation}
\frac{\partial \tilde{p}}{\partial X_j}-\lambda \frac{\partial D}{\partial X_j}=0, \qquad j=1,\ldots,m ,  \qquad  D=0.
\label{Dmodifiedextremal}
\end{equation}
Only those solutions of equations (\ref{Dmodifiedextremal}) that satisfy the condition (\ref{squeezingconstraint}), i.e. where $\mu^2$ is the maximum squeezing (and not, for example, the minimum one) are accepted and considered as potential optimal solutions. 
If it becomes necessary to impose a stronger constraint that two or more eigenvalues of $AA^\dagger$ are equal to $\mu^2$, then one can consider additional coefficients of the shifted characteristic polynomial
\begin{equation}
P(y)=\det\left[ (y+\mu^2)I -A A^\dagger\right]=\det\left[ (y+\mu^2)I -BK B^\dagger K\right],
\end{equation}
which can be expanded as 
\begin{equation}
P(y)=\sum_{j=0}^m D_j y^j.
\end{equation}
The coefficients $D_j$ are polynomial functions of $X_k$ and $D_0=D$. One can ensure that at least $l$ eigenvalues of $AA^\dagger$ are equal to $\mu^2$ by imposing the constraints 
\begin{equation}
D_0=D_1=\ldots,D_{l-1}=0.
\end{equation}

Let us finally consider the situation when the stellar rank $N$ of the conditionally generated state $|\psi_N\rangle$ \tcr{is so low} that not all the elements of matrix $B$ are determined by $|\psi_N\rangle$. In such case we can treat the conditions $d_k=\sqrt{N!/k!}c_k/c_N$  as constraints, where
 $d_k$ are the polynomial functions of $s_j$ and $\nu_{jk}$ defined in equation (\ref{dkdefinition}), and $c_k$ are the required complex amplitudes of the  heralded state $|\psi_N\rangle$ in the Fock basis. The constraints $d_k=\sqrt{N!/k!}c_k/c_N$ can be incorporated into the optimization of $p_S$ via Lagrange multipliers. To be specific, we shall consider even $N$, the case of odd $N$ can be treated completely analogously. For even $N$ the function that should be optimized can be expressed as $\tilde{p}-\sum_{l=0}^{N/2-1}\lambda_{l} d_{2l}-\sum_{l=0}^{N/2-1} \lambda_{l}^\ast d_{2l}^\ast $. Since $d_{2l}$ do not depend on $X_j$, the extremal equations (\ref{Xjextremal}) remain unchanged but are supplemented by the following additional extremal equations
\begin{equation}
\frac{\partial \tilde{p}}{\partial s_j}-\sum_{l=0}^{\frac{N}{2}-1} \lambda_l\frac{  \partial d_{2l}}{\partial s_j}=0, \qquad 
\frac{\partial \tilde{p}}{\partial \nu_{jk}}- \sum_{l=0}^{\frac{N}{2}-1}\lambda_l\frac{  \partial d_{2l}}{\partial \nu_{jk}}=0.
\label{underdetermined}
\end{equation}
Here $s_j$ and $s_j^\ast$ (and also $\nu_{jk}$ and $\nu_{jk}^\ast$) should be formally treated as independent variables when calculating the derivatives. Since also complex conjugates of $s_j$  and $\nu_{jk}$ appear in the above equations, it may be necessary to treat the real and imaginary parts of these variables as new (real) variables to recover a system of polynomial equations that combines Eqs. (\ref{Xjextremal}), (\ref{dkck}) and (\ref{underdetermined}).

\begin{figure}[!t!]
\includegraphics[width=0.75\linewidth]{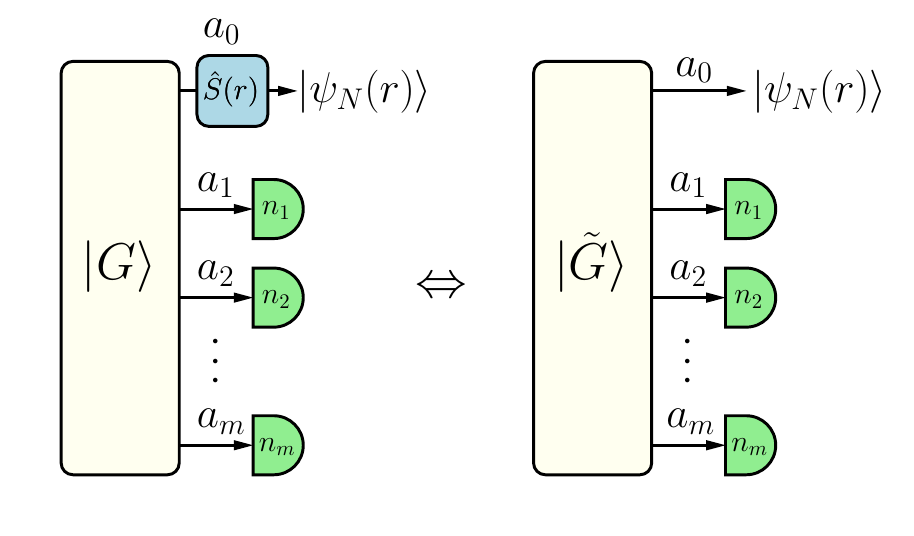}
\caption{Scheme for generation of single-mode squeezed superpositions of Fock states. The state generated from Gaussian core state $|G\rangle$ can be squeezed by unitary squeezing operation $\hat{S}(r)$ acting on mode $a_0$. This is equivalent to utilizing a different input multimode Gaussian state $|\tilde{G}\rangle$ with modified squeezing properties. 
}
\label{figsqueezed}
\end{figure}

\section{Generation of squeezed superpositions of Fock states}

It is often desirable to generate a squeezed superposition of  Fock states \cite{Fiurasek2005,Marek2008},
\begin{equation}
|\psi_N(r)\rangle= \hat{S}(r) \sum_{k=0}^N c_k|k\rangle,
\label{psiNsqueezed}
\end{equation}
instead of just the superposition of Fock states  (\ref{psidefinition}). Here
\begin{equation}
\hat{S}(r)=\exp\left[\frac{r}{2} (\hat{a}_0^{\dagger 2}-\hat{a}_0^2)\right]
\label{Srout}
\end{equation}
denotes the squeezing operator. To obtain the state (\ref{psiNsqueezed}), we can apply the squeezing operator $\hat{S}(r)$ to mode $a_0$ of the Gaussian core state $|G\rangle$ that generates the  state $\sum_{k=0}^N c_k|k\rangle$. As illustrated in Fig.~\ref{figsqueezed}, this additional squeezing can be incorporated into the preparation of the input Gaussian state, resulting in Gaussian state 
$|\tilde{G}\rangle$ whose covariance matrix can be easily calculated from the covariance matrix of $|G\rangle$ by applying the squeezing operation (\ref{Srout}) to mode $a_0$. The squeezing operation (\ref{Srout}) is unitary and deterministic hence it does not influence the heralding probability $p_S$ of state (\ref{psiNsqueezed}). However, it affects the squeezing of the input Gaussian state and it may happen that the maximum squeezing present in state $|\tilde{G}\rangle$ will be higher than the maximal squeezing in state $|G\rangle$.

If the goal is to maximize the heralding probability for limited available squeezing, we should take into account the required squeezing of the output state during  the maximization of $p_S$.
This can be achieved by incorporating the required output squeezing into the matrices $A$ and $B$. Specifically, we set
\begin{equation}
A_{00}=B_{00}=\tanh r,
\end{equation}
which takes us outside the class of the Gaussian core states. The conditionally generated state of mode $a_0$ can be expressed as 
\begin{equation}
|\psi_N(r)\rangle\propto \exp\left(\frac{\tanh r}{2}\hat{a}_0^{\dagger 2}\right) \sum_{k=0}^N d_k \hat{a}_0^{\dagger k}   |0\rangle.
\label{psiNsqueezedformula}
\end{equation}
The coefficients $d_k$ are  given by an equation fully analogous to (\ref{dkdefinition}) \tcr{where just have to take care of the nonzero $B_{00}$. Explicitly, we have}
\begin{equation}
\tcr{d_k=\left.\frac{1}{k!} \frac{\partial ^k}{\partial z_{0}^k} \prod_{j=1}^m \frac{\partial^{n_j}}{\partial z_j^{n_j}}\left[ \exp\left(-\frac{1}{2}B_{00}z_0^2\right)\exp \left(\frac{1}{2}\sum_{u=0}^m\sum_{v=0}^m B_{uv}z_u z_v\right) \right]\right|_{z_0=z_1=\ldots z_m=0}.}
\label{dkdefinitionsqueezed}
\end{equation}

  We can equivalently rewrite the state (\ref{psiNsqueezedformula}) as 
\begin{equation}
|\psi_N(r)\rangle\propto\sqrt{ \cosh r } \sum_{k=0}^N d_k \hat{a}_0^{\dagger k}  \hat{S}(r)|0\rangle= 
\sqrt{\cosh r}\hat{S}(r) \sum_{k=0}^N d_k\left (\hat{a}_0^\dagger\cosh r +\hat{a}_0\sinh r \right)^k |0\rangle.
\label{psiNrewritten}
\end{equation}
With the help of the identity
\begin{equation}
\left
(\hat{a}_0^\dagger\cosh r +\hat{a}_0\sinh r\right)^k |0\rangle =\sum_{j=0}^{k/2} \frac{k!}{2^j j! (k-2j)!}(\cosh r)^{k-j}(\sinh r)^j \hat{a}_0^{\dagger k-2j}|0\rangle
\end{equation}
we can establish system of linear equations for the  coefficients $d_k$,
\begin{equation}
\sum_{j=0}^{\frac{N-k}{2}} \frac{(k+2j)!}{2^j k! j!} (\cosh r)^{k+j-N} (\sinh r)^j  d_{k+2j} =\sqrt{\frac{N!}{k!}} \frac{c_k}{c_N},
\label{dksqueezing}
\end{equation}
which generalizes equation (\ref{dkck}). The system of linear equations (\ref{dksqueezing}) is in a triangular form. Therefore, it can be readily solved.
\tcr{Observe that the condition $d_N=1$  holds and remains valid for arbitrary output squeezing $r$. }

 The maximization of the heralding probability for the state (\ref{psiNsqueezed}) under the constraint of limited squeezing proceeds as follows. \tcr{First we solve the linear system of equations (\ref{dksqueezing}) to determine $d_k$. Next we have to  solve the system of polynomial equations (\ref{dkdefinitionsqueezed}) to determine all possible sets of the parameters $s_j$  and $\nu_{jk}$. From this point on, the machinery for determination of the maximum heralding probability $p_S$ developed in Section II applies directly, because  the derivation of the extremal equations for the damping parameters $X_j$ does not assume any specific form of the matrix $B$. Therefore, we have to construct and solve the extremal equations (\ref{Dmodifiedextremal}) which  will have the form of system of polynomial equations for $X_j$ and $\lambda$. The only difference is that the formula (\ref{pSdefinition}) for the heralding probability $p_S$ must be multiplied by a factor $(\cosh r)^{2N+1}$, which accounts for the squeezing of the conditionally generated state and follows from equation (\ref{psiNrewritten}). }

\section{Two heralding modes}
In this section we apply the general procedure to a three-mode configuration with signal mode $a_0$ and two heralding modes $a_1$ and $a_2$. The
three-mode Gaussian core state can be expressed as 
\begin{equation}
|G\rangle=Z^{1/4}\exp\left( x_1\hat{a}_0^\dagger \hat{a}_1^\dagger+x_2\hat{a}_0^\dagger \hat{a}_2^\dagger+x_1x_2\nu \hat{a}_1^\dagger \hat{a}_2^\dagger +\frac{s_1 x_1^2}{2}\hat{a}_1^{\dagger 2}+ \frac{s_2 x_2^2}{2}\hat{a}_2^{\dagger 2}  \right) |0,0,0\rangle.
\end{equation}
The corresponding matrices $B$ and $K$ read
\begin{equation}
B=\left(
\begin{array}{ccc}
0 & 1 & 1 \\
1 &  s_1 &  \nu \\
1 &  \nu &  s_2
\end{array}
\right),
\qquad 
K=\left(
\begin{array}{ccc}
1 & 0 & 0 \\
0 &  X_1 &  0 \\
0 & 0 & X_2
\end{array}
\right).
\end{equation}
For brevity, we have dropped the subscript $12$ in the off-diagonal matrix element $\nu$.
We shall consider generation of even superpositions of Fock states up to Fock number $N=6$ and generation of odd superpositions of Fock states up to Fock number $N=7$. For these target states, the number of parameters of the matrix $B$ exactly matches the number of independent equations for state coefficients $d_k$. 

Let us first focus on the generation of even superpositions of Fock states,
\begin{equation}
|\psi_6\rangle=\sum_{k=0}^3 c_{2k} |2k\rangle.
\label{psi6definition}
\end{equation}
There are at least two detection patterns that can lead to preparation of this state. The first pattern is symmetric, $(n_1,n_2)=(3,3)$. Using equation (\ref{dkdefinition}) we obtain the following expressions for the state coefficients $d_{2k}$ corresponding to even Fock states ($d_{2k+1}$ vanish since the total number of counted photons is even):
\begin{eqnarray}
d_0&=&9 s_1 s_2 \nu+ 6 \nu^3, \nonumber \\
d_2&=&9 \left[s_1s_2 +(s_1+s_2)\nu + 2\nu^2\right], \nonumber \\
d_4&=& 3 (s_1 + s_2 + 3 \nu), \nonumber \\
d_6&=&1.
\label{deven33}
\end{eqnarray}
The above formulas constitute a system of equations for parameters $s_j$ and $\nu$. Note that similar formulas for the parameters of conditionally generated states in the Gaussian boson sampling setup were derived for lower numbers of total counted photons in Ref. \cite{Su2019}. The present parameters  $s_1$, $s_2$, and $\nu$ correspond to parameters $f_{22}^\ast$, $f_{33}^\ast$, and $f_{23}^\ast$ in Ref. \cite{Su2019}.  Owing to the symmetry of the expressions in equation (\ref{deven33}), the variables $s_1$ and $s_2$ can be readily eliminated, and after some algebra one arrives at a cubic equation for  $\nu$,
\begin{equation}
  15 \nu^3 - 3 d_4 \nu^2 + d_2 \nu-d_0=0.
\end{equation}
For each root $\nu$,  the other two parameters $s_1$ and $s_2$ are given by
\begin{eqnarray}
s_1 &=& \frac{1}{6}\left(d_4 - 9 \nu \pm \sqrt{d_4^2 -4 d_2  - 6 d_4 \nu + 45 \nu^2}\right), \nonumber \\ 
 s_2 &=& \frac{1}{6}\left(d_4 - 9\nu \mp \sqrt{d_4^2-4 d_2  - 6 d_4 \nu + 45 \nu^2}\right).
\end{eqnarray}
Since the detection pattern is symmetric, the two solutions for $s_1$ and $s_2$ are related by a simple permutation $s_1 \leftrightarrow s_2$. 

We must also consider the asymmetric detection pattern $(n_1,n_2)=(4,2)$. Following the same steps as before we obtain
\begin{eqnarray}
d_0&=& 3 s_1 (s_1 s_2 + 4 \nu^2),  \nonumber  \\
d_2&=& 3 (s_1^2 + 4 \nu^2 + 2 s_1 (s_2 + 4 \nu)), \nonumber \\
d_4&=& 6 s_1 + s_2 + 8 \nu, \nonumber \\
d_6&=&1.
\label{dk42}
\end{eqnarray}
This system of polynomial equations can be converted into a polynomial equation for a single variable $\nu$ via construction of the Gröbner basis, which yields
\begin{eqnarray}
& &-1331 d_0^2 - 12 d_2^3 + 154 d_0 d_2 d_4 + d_2^2 d_4^2 - 12 d_0 d_4^3 + 
 352 d_0 d_2 \nu + 32 d_2^2 d_4 \nu - 384 d_0 d_4^2 \nu - 416 d_2^2 \nu^2  \nonumber  \\
& & +   4704 d_0 d_4 \nu^2 - 23520 d_0 \nu^3 + 96 d_2 d_4 \nu^3 + 5520 d_2 \nu^4 - 
 144 d_4^2 \nu^4 - 5760 d_4 \nu^5 + 14400 \nu^6=0. \nonumber \\
 \label{nuformula42}
\end{eqnarray}
Note that the resulting polynomial has a degree $6$ and is much more complicated than for the  detection pattern $(3,3)$, despite the same total number of detected photons. 
Once the roots $\nu$ are found, the parameters $s_1$ and $s_2$ can be determined from the formulas
\begin{eqnarray}
& & s_1 =\frac{1}{33}\left(3 d_4 - 12 \nu \mp \sqrt{3} \sqrt{ 3 d_4^2 -11 d_2 - 24 d_4 \nu + 180 \nu^2}\right), \nonumber \\ 
& &    s_2=\frac{1}{11} \left (5 d_4 - 64 \nu  \pm 2 \sqrt{3} \sqrt{3 d_4^2 -11 d_2 - 24 d_4 \nu + 180 \nu^2}\right). 
\label{s12pattern42}
\end{eqnarray}
Here, the permutation symmetry is broken because the detection pattern is asymmetric.  Bézout's theorem states that if the number of roots is finite, then the maximum number of  roots of system of $m$ polynomial equations for $m$ variables is the product of the degrees of the polynomials.  Consequently, we can expect at most  $6=3\times 2\times 1$ different solutions of the system of Eqs. (\ref{dk42}).  Therefore, some of the roots of equation (\ref{nuformula42}) combined with equation (\ref{s12pattern42})
may not represent a  solution of the original system of Eqs. (\ref{dk42}) and each potential solution must be checked whether it actually solves the system (\ref{dk42}). In fact, using the Gröbner basis construction,  it is possible to uniquely express $s_1$ and $s_2$ as functions of $\nu$  for each solution $\nu$ of equation (\ref{nuformula42}). However, the resulting expressions are very lengthy \tcr{and we} therefore present the potential solutions $s_1$  and $s_2$ in a more compact form as roots of quadratic equations (\ref{s12pattern42}).

Note that the heralding pattern $(5,1)$ is generally not suitable for preparation of an arbitrary state of the form (\ref{psi6definition}), because for this pattern the state coefficients $d_{2k}$ do not depend on $s_2$ and the number of parameters that can be tuned to engineer the state as required is reduced from three to two.

\begin{figure}[!t!]
\centering\includegraphics[width=0.8\linewidth]{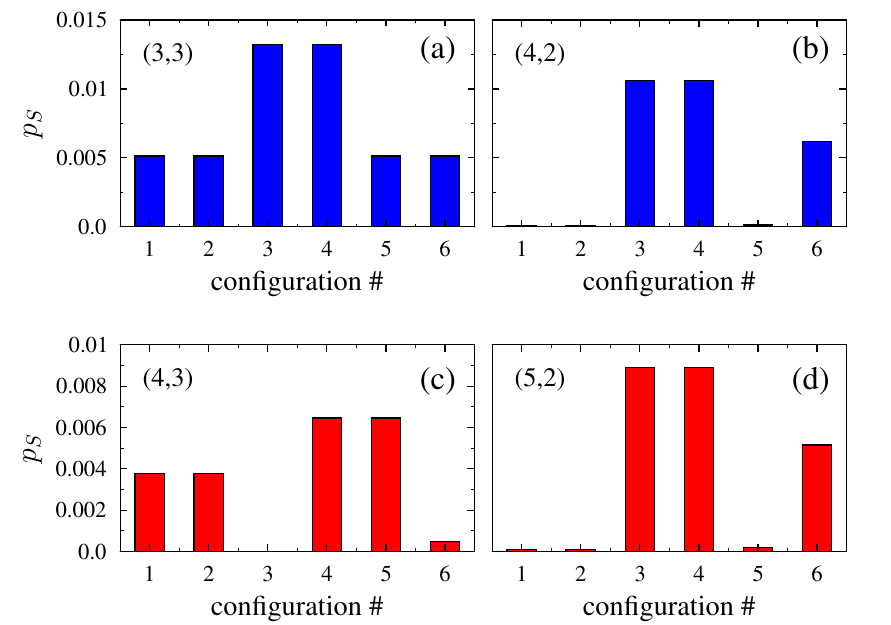}
\caption{Results of maximization of the heralding probability $p_S$ for balanced superposition target states $|\psi_6\rangle$ (a,b) and $|\psi_7\rangle$ (c,d) defined in equation (\ref{psi67specific}). For each  heralding pattern $(n_1,n_2)$ considered, six configurations $(s_1,s_2,\nu)$ generating the state were identified and the maximum achievable heralding probability $p_S$ for each configuration is plotted. Note that all the plotted $p_S$ are nonzero, although some of them are very small. See also Table I in the Appendix B.} 
\label{figtwo}
\end{figure}

\begin{figure}[!t!]
\centering\includegraphics[width=0.8\linewidth]{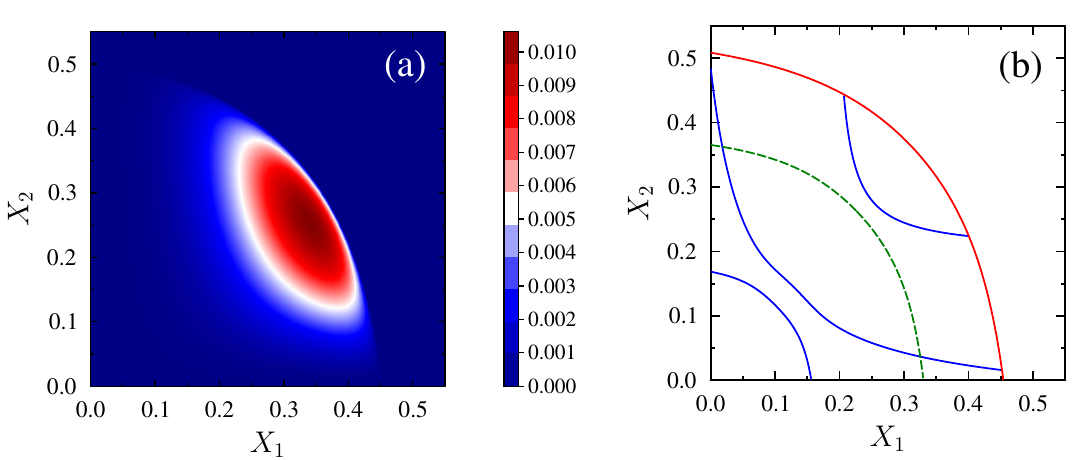}
\caption{(a) Dependence of the heralding probability $p_S$ on $X_1$ and $X_2$ is plotted for target state $|\psi_6\rangle$, heralding pattern $(4,2)$, and the optimal configuration 
$s_1=-0.072-1.199i$, $s_2=-0.960-0.109i$, $\nu=0.859+0.913i$. The set of points satisfying the squeezing constraint $D=0$ is plotted for this configuration in panel (b) for two different squeezing values $\mu^2=0.25$ (blue solid lines) and $\mu^2=0.65$ (green dashed line). The red curve represents the physical boundary in the $(X_1,X_2)$ plane where the largest eigenvalue of $AA^\dagger$ becomes equal to $1$. }
\label{figsc}
\end{figure}

Next, we  give an example of the  generation of superposition of odd Fock states. Specifically, we consider the preparation of the  state
\begin{equation}
|\psi_7\rangle=\sum_{k=0}^3 c_{2k+1}|2k+1\rangle.
\label{psi7definition}
\end{equation}
We list here the results for the  detection pattern $(n_1,n_2)=(4,3)$. The state coefficients read 
\begin{eqnarray}
d_1&=& 9 s_1^2 s_2 + 24 \nu^3 + 36 s_1 \nu (s_2 + \nu), \nonumber \\
d_3&=& 3 (s_1^2 + 6 s_1 (s_2 + 2 \nu) + 4 \nu (s_2 + 3\nu)), \nonumber \\
d_5&=& 3 (2 s_1 + s_2 + 4 \nu), \nonumber \\ 
d_7&=&1.
\label{dkpattern43}
\end{eqnarray}
The parameters $s_1$ and $s_2$ can be analytically eliminated which yields polynomial equation for $\nu$,
\begin{eqnarray}
& & -1331 d_1^2 - 12 d_3^3 + 154 d_1 d_3 d_5 + d_3^2 d_5^2 - 12 d_1 d_5^3 + 
 4752 d_1 d_3 \nu - 144 d_3^2 d_5 \nu - 784 d_1 d_5^2 \nu  \nonumber \\
 & &  + 16 d_3 d_5^3 \nu - 5616 d_3^2 \nu^2 - 7296 d_1 d_5 \nu^2 + 2112 d_3 d_5^2 \nu^2 - 80 d_5^4 \nu^2 + 
 51072 d_1 \nu^3 + 15168 d_3 d_5 \nu^3  \nonumber \\
& & - 4800 d_5^3 \nu^3  - 117936 d_3 \nu^4 + 
 37680 d_5^2 \nu^4 - 80640 d_5 \nu^5 + 282240 \nu^6=0. \nonumber \\
 \label{nuequation43}
\end{eqnarray}
Note that the degree of this polynomial equation, as well as the degrees of the polynomials in equation (\ref{dkpattern43}) is the same as that for the detection pattern $(3,3)$, although  the total number of detected photons increased to $7$. Similarly as before, once the candidate optimal $\nu$ are calculated as roots of (\ref{nuequation43}), the corresponding  $s_1$ and $s_2$ can be obtained by solving quadratic equation, which yields
\begin{eqnarray}
& & s_1=\frac{1}{33} \left(3 d_5 - 30 \nu \pm  \sqrt{3} \sqrt{3 d_5^2-11 d_3  - 16 d_5 \nu + 168 \nu^2}\right), \nonumber \\
& &  s_2 = \frac{1}{33}\left (5 d_5 - 72 \nu \mp  2 \sqrt{3} \sqrt{ 3 d_5^2 -11 d_3 - 16 d_5 \nu + 168 \nu^2}\right).
\end{eqnarray}
Just as  for the heralding pattern $(4,2)$, each candidate solution must be checked whether it actually solves the original system of equations (\ref{dkpattern43}).

Let us now turn attention to  the optimization of the preparation probability $p_S$ of the non-Gaussian states (\ref{psi6definition}) and (\ref{psi7definition}). For the considered two-mode heralding patterns, all parameters of matrix $B$ are fixed by the structure of the targeted state (up to the existence of multiple solutions to the polynomial equations).  There thus remain the two  variables $X_1$  and $X_2$ that influence only $p_S$ and can be optimized. The polynomial  equations (\ref{Qextremal}) for the optimal \tcr{$X_1$ and $X_2$} can be explicitly written as 
\begin{equation}
Q_1=\sum_{j=0}^2 \sum_{k=0}^2 q_{jk}^{(1)} X_1^j X_2^k=0, \qquad Q_2=\sum_{j=0}^2 \sum_{k=0}^2 q_{jk}^{(2)} X_1^j X_2^k=0.
\label{Qextremaltwomode}
\end{equation}
\tcr{In this formula $j$ and $k$ denote the powers of the variables $X_1$ and $X_2$, respectively.}
Explicit expressions for the  coefficients $q_{jk}^{(l)}$ are provided in the Appendix A.

We illustrate the optimization of probability $p_S$ for the preparation of balanced superpositions of even or odd Fock states,
\begin{equation}
|\psi_6\rangle=\frac{1}{2}\left(|0\rangle+|2\rangle+|4\rangle+|6\right), \qquad
|\psi_7\rangle=\frac{1}{2}\left(|1\rangle+|3\rangle+|5\rangle+|7\right).
\label{psi67specific}
\end{equation}
For  target state $|\psi_6\rangle$ we consider the heralding patterns $(3,3)$ and $(4,2)$, and for target state $|\psi_7\rangle$ we consider patterns $(4,3)$ and $(5,2)$. For each heralding pattern, we utilized the formulas given above to determine all the sets of parameters $(s_1,s_2,\nu)$ that generate the state $|\psi_6\rangle$ (or $|\psi_7\rangle$) upon successful heralding. For each set of parameters  $(s_1,s_2,\nu)$ and $(n_1,n_2)$  we then numerically solve the system of two polynomial equations  (\ref{Qextremaltwomode}) to identify the potentially optimal parameters $X_1$ and $X_2$. Note that only positive real solutions are acceptable, and for each candidate solution we must check whether the physicality condition $I-AA^\dagger >0$ is satisfied. For all valid solutions $X_1$ and $X_2$ we evaluate $p_S$ and finally we take the maximum of $p_S$ over all the configurations found. The parameters of all the configurations are listed in the Appendix B.
The systems of polynomial equations were solved in Wolfram Mathematica using the functions \texttt{GroebnerBasis}, \texttt{Solve}, and \texttt{NSolve}. \tcr{Note that the function \texttt{NSolve}  employs the homotopy continuation method to numerically calcuate all isolated roots of a system of polynomial equations.} 

For each heralding pattern $(n_1,n_2)$ we have identified  six configurations $(s_1,s_2,\nu)$ that generate the required target state. In Fig.~\ref{figtwo} we plot the maximum heralding probability $p_s$ for each such configuration. Since some of these configurations differ only by complex conjugation, we can have pairs of configurations with identical  $p_S$. Moreover for the symmetric heralding pattern $(3,3)$ the permutation symmetry $s_1\leftrightarrow s_2$ induces further degeneracy. For $N=6$ the symmetric heralding pattern $(3,3)$ is the optimal choice for the considered target state $|\psi_6\rangle$. However, it is easy to find an example of target state for which the asymmetric pattern $(4,2)$ is optimal. Therefore, it is important to consider all possible heralding patterns. We have verified by exhaustive search in the plane $(X_1,X_2)$ that the calculated heralding probabilities are indeed the global maxima for each configuration. In fact, for all the $24$ configurations  considered, the heralding probability $p_S$ exhibits only a single maximum in the region of physically allowed pairs $(X_1,X_2)$. An example of the dependence of $p_S$ on $X_1$ and $X_2$ is shown in Fig.~\ref{figsc}(a). \tcr{It is an open question whether the heralding probability $p_S$ can exhibit multiple extrema in the domain of physically allowed values of $X_j$.}

\begin{figure}[!t!]
\centering\includegraphics[width=0.5\linewidth]{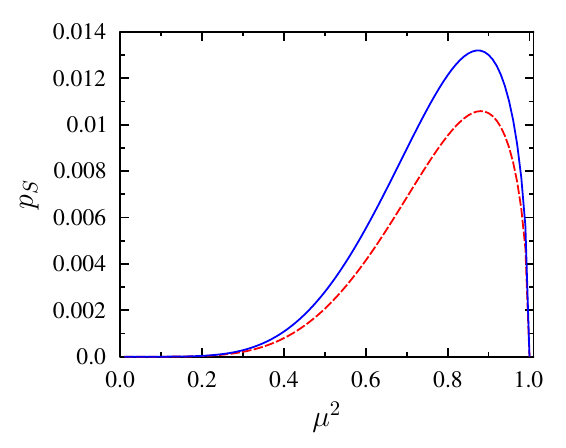}
\caption{Maximum heralding probability for target state $|\psi_6\rangle$ defined in equation (\ref{psi67specific}) when the maximum squeezing is set to $\mu^2=\tanh ^2 r_{\mathrm{max}}$. The results are plotted for two heralding patterns (3,3) (blue solid line) and (4,2) (red dashed line).}
\label{figpSmu}
\end{figure}

We now illustrate the maximization of the heralding probability $p_S$  under the constraint of limited available squeezing bounded by the maximum single-mode squeezing constant $r_\mathrm{max}$.  The determinant $D$ defined in equation (\ref{Ddefinition})  explicitly reads
\begin{eqnarray}
D&=&\mu^6-|s_1+s_2-2\nu|^2X_1^2X_2^2
-\mu^4  \left(|s_1|^2 X_1^2 + |s_2|^2  X_2^2 + 2( X_1+X_2) + 2|\nu|^2 X_1 X_2\right)  \nonumber \\
& & +  \mu^2 \left((X_1 +X_2)^2 +2|s_1-\nu|^2X_1^2X_2+2|s_2-\nu|^2X_1X_2^2+ |s_1s_2-\nu^2|^2 X_1^2X_2^2  \right).
\end{eqnarray}
The constraint $D=0$ restricts the optimization to curves in the plane $(X_1,X_2)$. Examples of  the curves implicitly defined by the equation $D=0$ are plotted in Fig.~\ref{figsc}(b) for two different values of $\mu^2$. In addition, the physical boundary corresponding to $\mu^2=1$ is displayed  there. In this specific example,   we obtain three separate curves for $\mu=0.25$ whereas we obtain  only a single curve for $\mu^2=0.65$.  The three different curves for  $\mu^2=0.25$ correspond to $\mu^2$ being the largest, medium, and  lowest eigenvalue of $AA^\dagger$, respectively.

For two heralding modes, the Lagrange multiplier $\lambda$ can be easily eliminated from Eqs. (\ref{Dmodifiedextremal}) and we arrive at the following set of two polynomial equations for $X_1$ and $X_2$,
\begin{equation}
\frac{\partial \tilde{p}}{\partial X_1}\frac{\partial D}{\partial X_2}=\frac{\partial \tilde{p}}{\partial X_2}\frac{\partial D}{\partial X_1},   \qquad D=0.
\end{equation}
For every solution of this system of polynomial equations we must  check whether $X_j>0$ and whether the largest eigenvalue of $K^{1/2}BKB^\dagger K^{1/2}$ does not exceed $\mu^2$. Only  solutions that satisfy these conditions are accepted. The outcome of the  constrained optimization procedure is plotted in Fig.~\ref{figpSmu} for the generation of state $|\psi_6\rangle$. The achievable heralding probability $p_S$ increases with increasing available squeezing $\mu^2$ until it reaches its global maximum. Then in begins to fall again because when we fix the squeezing to be too high we leave the optimal region. The maxima of the curves for the heralding patterns $(3,3)$ and $(4,2)$ correspond to the maxima of the heralding probabilities plotted in Fig.~2(a) and 2(b), respectively.

\tcr{\section{Three heralding modes}}

\tcr{In this section we present examples of optimization of the heralding probability $p_S$ for three heralding modes, $m=3$, a configuration realized in the recent landmark experiment by Xanadu \cite{Larsen2025}.  For  fixed values of $s_j$ and $\nu_{jk}$ and a symmetric heralding pattern $n_1=n_2=n_3=n$ we have optimized the heralding probability $p_S$  for several values of $n$ up to $n=5$, which corresponds to detection of $15$ heralding photons in total. We have chosen three different sets of values of $s_j$ and $\nu_{jk}$ to illustrate the different scenarios that can occur when solving the system of polynomial equations for the optimal $X_j$.  As a first example we consider real $\nu_{jk}$ and purely imaginary $s_j$: $\nu_{12}=\frac{6}{7}$, $\nu_{13}=-\frac{5}{11}$, $\nu_{23}=\frac{2}{3}$, and $s_1=s_2=s_3=i$. In this case the matrix $B$ is complex and the system of equations $Q_j=0$ has finite number of roots, as explicitly verified by  construction of the Gröbner basis. The resulting maximum achievable $p_S$ is plotted as a function of $n$ in Fig.~6(a). }

\tcr{We next give an example with real parameters and set $s_1=s_2=s_3=1$. The matrix $B$ becomes real and the system of polynomial equations $Q_j=0$, $j=1,2,3,$ possesses infinitely many solutions, as discussed in Sec. II. This can be remedied by constructing three additional polynomials (\ref{EFeliminated}) for three sets of indices $(ij,kl)=[(12,13),(12,23),(13,23)]$. For the symmetric heralding pattern $(n,n,n)$ these three additional polynomials coincide, 
\begin{equation}
 W_{12,13}=W_{13,23}=W_{12,23}.
\end{equation}
Therefore, it suffices to consider only one of them, say $W_{12,13}$. The system of four polynomial equations
\begin{equation}
Q_1=0,\qquad Q_2=0,\qquad Q_3=0, \qquad W_{12,13}=0,
\label{QWsystem}
\end{equation}
has a finite number of solutions, as confirmed by construction of the Gröbner basis, which yields a polynomial equation of degree $33$ for the variable $X_1$. We check all the solutions to find the optimal values of $X_j$ and plot the resulting maximum heralding probability $p_S$ in Fig. 6(b). }

\begin{figure}[t]
\centerline{\includegraphics[width=0.95\linewidth]{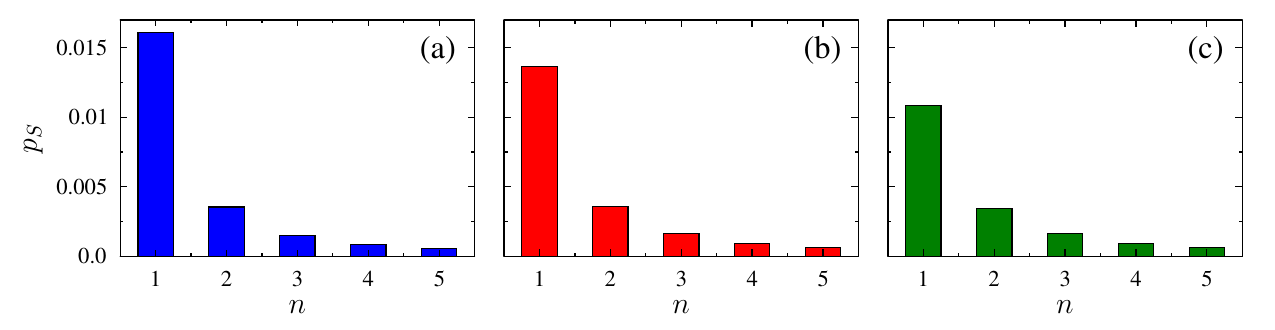}}
\caption{\tcr{Maximum heralding probabilities for a configuration with three heralding modes, symmetric heralding pattern $n_1=n_2=n_3=n$, and three sets of control parameters of the input Gaussian state $|G\rangle$: (a) complex matrix $B$ with $\nu_{12}=\frac{6}{7}$, $\nu_{13}=-\frac{5}{11}$, $\nu_{23}=\frac{2}{3}$, and $s_1=s_2=s_3=i$,  (b) real matrix $B$ with $s_1=s_2=s_3=1$, and (c) a highly symmetric configuration with $\nu_{12}=\nu_{13}=\nu_{23}=1$ and $s_1=s_2=s_3=2$.}}
\label{figthreeheraldingmodes}
\end{figure}

\tcr{Numerical simulations suggest that the inclusion of the polynomial $W_{12,13}$ ensures existence of a finite number of solutions for generic randomly chosen values of $s_j$ and $\nu_{jk}$. However, even the system of equations (\ref{QWsystem}) can possess infinitely many solutions  for specific and highly symmetric choice of the parameters $\nu_{jk}$ and $s_j$. To give an example, we consider the fully permutationally invariant set of parameters  $s_1=s_2=s_3=2$ and $\nu_{12}=\nu_{13}=\nu_{23}=1$.  The system of equations (\ref{QWsystem}) possesses infinitely many solutions. A closer inspection reveals that the polynomials $Q_j$ can be expressed as 
\begin{equation}
Q_1=(1+X_2)(1+X_3)\tilde{Q}_1,\qquad Q_2=(1+X_1)(1+X_3)\tilde{Q}_2,\qquad Q_3=(1+X_1)(1+X_2)\tilde{Q}_3,
\end{equation}
where $\tilde{Q}_j$ denote some polynomials, and  the additional polynomials $W_{ij,kl}$ read
\begin{equation}
W_{12,13}=W_{12,23}=W_{13,23}=8n^2(X_1-X_2)(X_1-X_3)(X_2-X_3).
\end{equation}
The infinite unphysical set of solutions thus has the form $X_1=-1$, $X_2=-1$, and $X_3$ arbitrary, and permutations thereof. We can remove these unphysical solutions by considering the system of equations $\tilde{Q}_j=0$ which has a finite number of solutions. Note also that the polynomials $W_{ij,kl}$ provide a useful information because they indicate that the optimal solution must exhibit some symmetry. Indeed, we find that  the optimal solution satisfies  $X_1=X_2=X_3$ for all the $n$ considered. The corresponding maximum heralding probabilities $p_S$ are plotted in Fig.~6(c).}  

\tcr{Analysis of the dependence of $p_S$ on $n$ plotted in Fig.~6 suggests that the scaling is sub-exponential and rather resembles an inverse  polynomial dependence \cite{Fiurasek2025arXiv}, $p_S \propto n^{-\gamma}$ with $\gamma$ close to $2$. Achieving  rather large number of heralding photons is thus in principle experimentally feasible \cite{Larsen2025}. In the present example, heralding probabilities of the order of $5\times 10^{-4}$ are obtained  for $15$ heralding photons. As shown in Refs. \cite{Larsen2025,Hanamura2025,Endo2026}, increasing the number of heralding photons for a fixed number of heralding modes can help to improve the properties of the generated states, such as the nonlinear squeezing or the effective amplitude of the generated approximate superposition of coherent states, without significantly increasing the complexity of the experiment. }

\tcr{Solving the system of polynomial equations for $X_j$ with Mathematica is fast for $m=2$ and $m=3$  and the calculation takes only a fraction of a second for $m=2$ and less than two seconds for $m=3$, respectively, on a standard desktop machine. However, the complexity and the number of solutions increases quickly with $m$. The degree of each polynomial $Q_j$ is $2m$ but the only monomial with this degree, $\prod_{j=1}^m X_j^2$, is the same for all $Q_j$. Therefore, the degrees of $m-1$ polynomials can be readily reduced to $2m-1$. Bézout's bound then predicts that the system can have up to $2m(2m-1)^{m-1}$ roots. For $m=4$ the calculations are still feasible on a desktop computer, but it takes about $8$ minutes to find the solutions with \texttt{NSolve}. Moreover, for $m=4$ the system of equations $Q_j=0$  can exhibit infinitely many solutions even for complex $B$ and the degeneracy must be broken, e.g.,  by making a tiny modification of the function $p_S$, similarly as in equation (\ref{pSepsilon}). On the other hand, the complexity of the system of equations for the optimal $X_j$ does not increase with the total number o heralding photons for a fixed number of heralding modes.}

\tcr{With $m$ heralding modes the number of the control parameters of the input Gaussian state $|G\rangle$ that specify the matrix $B$ and can be used to design the output state $|\psi_N\rangle$ scales quadratically with $m$, $\mathcal{M}=\frac{1}{2}m(m+1)$.  In principle, it could thus be possible to engineer arbitrary finite superposition of even Fock states up to $N=m(m+1)$, or arbitrary superposition of odd Fock states up to $N=m(m+1)+1$. However, the complexity of the system of equations for the parameters of matrix $B$ increases quickly and Bézout's bound predicts that it can have up to $\mathcal{M}!$ roots. This rises from $6$ roots for $m=2$ to $720$ for $m=3$. }

\tcr{Additionally, the number of the possible heralding patterns also increases. We can require that at least two photons are detected in each heralding mode, $n_j \geq 2$, to ensure that the amplitudes $c_k$ of the generated state depend on $s_j$. 
The number of the heralding patterns is then equal to the number of ways to express the integer $N-2m$ as a sum of at most $m$ integers. To be  specific, let us consider generation of superpositions of even Fock states up to $N=m(m+1)$.  As discussed in Section IV, there exist two different heralding patterns for $m=2$. This increases to $7$ heralding patterns for $m=3$, and $34$ patterns for $m=4$.  Generally, the optimization should be performed over all possible heralding patterns. }

\bigskip

\section{Single-rail entangled qubits}

In this section we apply the formalism to generation of two-mode entangled state 
\begin{equation}
|\Phi(w)\rangle=\frac{1}{\sqrt{1+w^2}}(|11\rangle+w|00\rangle),
\label{Phitwomode}
\end{equation}
where  $w$ is a real parameter. This state can be interpreted as entangled state of two single-rail photonic qubits encoded into superpositions of vacuum and single-photon states \cite{Ralphsinglerail}. \tcr{Generation of the state (\ref{Phitwomode}) is closely related to generation of heralded single-photon states, correlated photon pairs, and heralded entangled states of two photons \cite{Castelletto2008,Wagenknecht2010,Scott2020,Adam2024,Marcellino2024,Forbes2025,Huang2025}. However, for $w\neq 0$ the two-mode state (\ref{Phitwomode}) is not an eigenstate of the total photon number operator.}

\begin{figure}[t]
\centerline{\includegraphics[width=0.47\linewidth]{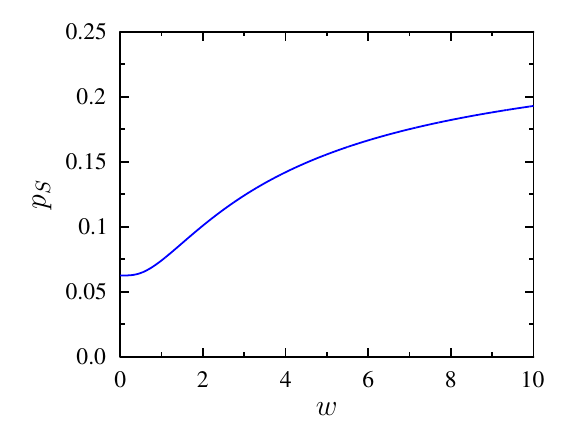}}
\caption{Optimal heralding probability $p_S$ of generation of the entangled two-mode state $|\Phi(w)\rangle$ is plotted as function of the state parameter $w$.}
\label{figsinglerail}
\end{figure}

Since the stellar rank of the state (\ref{Phitwomode}) is $2$, detection of two heralding photons is required to generate it from input Gaussian state. It is easy to see that a single heralding mode is \tcr{not sufficient} and two heralding modes with heralding pattern $(1,1)$ are required. Therefore, the source Gaussian \tcr{state is} a four-mode state that is divided into two signal modes and two heralding modes. The corresponding matrices $B$ and $K$ which specify the state read
\begin{equation}
B=\left(
\begin{array}{cccc}
 0 & 0 & 1 & 0 \\
 0 & 0 & 0 & 1 \\
 1 & 0 & s_1 & w \\
 0 & 1 & w & s_2 
\end{array}
\right), 
\qquad
K=\left(
\begin{array}{cccc}
1 & 0 & 0 & 0 \\
 0 &1 & 0 & 0 \\
0 & 0 & X_1 & 0 \\
 0 & 0 & 0  & X_2 
\end{array}
\right). 
\end{equation}
If we associate creation operators $\hat{a}^\dagger$ and  $\hat{b}^\dagger$ with the signal modes and $\hat{c}_1^\dagger$ and $\hat{c}_2^\dagger$ with the heralding modes, we can explicitly express the source four-mode Gaussian state as 
\begin{equation}
|G\rangle=Z^{1/4} \exp\left(x_1 \hat{a}^\dagger \hat{c}_1^\dagger +x_2 \hat{b}^\dagger \hat{c}_2^\dagger +x_1x_2w \hat{c}_1^\dagger \hat{c}_2^\dagger+\frac{s_1 x_1^2}{2}\hat{c}_1^{\dagger 2}+\frac{s_2 x_2^2}{2}\hat{c}_2^{\dagger 2}\right)|\mathrm{vac}\rangle.
\end{equation}
Recall that $X_j=x_j^2$. 
While the off-diagonal elements of matrix $B$ are fixed by the structure of the heralded state (\ref{Phitwomode}), the diagonal elements $s_1$ and $s_2$ are free parameters that influence only the probability $p_S$ of successful heralding,
\begin{equation}
p_S=\left(1+w^2\right)X_1 X_2\left[\det(I-BKB^\dagger K)\right]^{1/2}.
\label{pStwomode}
\end{equation}
 Let us first assume that $s_1=s_2=0$. The extremal equations 
 \begin{equation}
 \frac{\partial p_S}{\partial X_1}=0, \qquad \frac{\partial p_S}{ \partial X_2}=0,
 \label{Xextremaltwomode}
 \end{equation}
can be solved analytically, and the optimal damping parameters $X_1$ and $X_2$ read
\begin{equation}
X_1=X_2=\frac{3-\sqrt{1+8w^2}}{4(1-w^2)}.
\end{equation}
Upon inserting these expressions into equation (\ref{pStwomode}), we obtain formula for the heralding probability
\begin{equation}
p_S=(1+w^2) \frac{\left(3-\sqrt{1+8w^2}\right)^2}{128 (1-w^2)^3}\left[1-4w^2+\sqrt{1+8w^2}\right].
\label{pSsinglerailanalytical}
\end{equation}
The heralding probability (\ref{pSsinglerailanalytical}) is plotted in Fig.~\ref{figsinglerail}. Note that $p_S$ \tcr{ is a monotonically} increasing function of $w$. At $w=0$ the state (\ref{Phitwomode}) becomes the two-photon Fock state $|1,1\rangle$ and $p_S$ becomes equal to the square of the maximum heralding probability  of Fock state $|1\rangle$, $p_S=1/16=(1/4)^2$ \tcr{\cite{Christ2012}}. In the limit $w\rightarrow\infty$ we obtain  $p_S=1/4$, which corresponds to the maximum fidelity of the state $|1,1\rangle$ with a Gaussian state \cite{Chabaud2021}. Note that although the generated state (\ref{Phitwomode}) becomes a Gaussian vacuum state in the limit $w\rightarrow \infty$, the heralding probability is strictly less than one because the two heralding photons must be generated  to trigger the detectors.

Let us now investigate whether nonzero  parameters $s_1$ and $s_2$ can further increase the heralding probability (\ref{pStwomode}). Note that only the determinant $D=\det(I-BKB^\dagger K)$ in equation (\ref{pStwomode}) depends on $s_1$ and $s_2$.  Analysis of the analytical formula for $D$ reveals that the only term dependent on the complex phases of $s_1$ and $s_2$ is proportional to $s_1s_2+s_1^\ast s_2^\ast$,
\begin{equation}
D=\tilde{D}(|s_1|,|s_2|,X_1,X_2)-(s_1s_2+s_1^\ast s_2^\ast)w^2X_1^2X_2^2.
\label{Dsinglerail}
\end{equation}
 Since our goal is to maximize the determinant, we can, without loss of generality, assume real $s_1$ and $s_2$, which maximizes the determinant when the parameters $s_1$ and $s_2$ have opposite signs. To show this, consider any set of parameters $(s_1,s_2,X_1,X_2)$ that satisfy the condition $I-K^{1/2}BKB^\dagger K^{1/2}>0$. We can continuously change the phases of complex $s_1$ and $s_2$ such that finally $s_1$ and $s_2$ become real with opposite signs, e.g., $s_1^\prime=|s_1|$ and $s_2^\prime=-|s_2|$. The phases can be continuously changed such that the value of the determinant (\ref{Dsinglerail}) does not decrease during this phase tuning. Since the set of eigenvalues of a Hermitian matrix depends continuously on the matrix parameters and since the determinant is the product of eigenvalues we can conclude that the replacement of $s_1$ and $s_2$ by $s_1^\prime$ and $s_2^\prime$ preserves the physicality of the configuration, that is, all eigenvalues of  $I-K^{1/2}BKB^\dagger K^{1/2}$ remain positive.

When $s_1$ and $s_2$ are real, then $B=B^\dagger$ and the determinant in equation (\ref{pStwomode}) factorizes,
\begin{equation}
\tilde{p}=p_S^2=\left(1+w^2\right)^2X_1^2 X_2^2 \det(I-BK)\det(I+BK).
\end{equation}
 The potentially optimal values of $X_1$, $X_2$, $s_1$, and $s_2$ should satisfy the extremality conditions
\begin{equation}
\frac{\partial \tilde{p}}{\partial X_1}=0, \qquad \frac{\partial \tilde{p}}{\partial X_2}=0, \qquad \frac{\partial \tilde{p}}{\partial s_1}=0, \qquad \frac{\partial \tilde{p}}{\partial s_2}=0.
\label{ptildeextremaltwomode}
\end{equation}
This system of polynomial equations for $s_1$, $s_2$, $X_1$, and $X_2$ has infinitely many solutions, because any set of these variables which solves the two equations $\det(I-BK)=0$ and $\det(I+BK)=0$ also solves the system (\ref{ptildeextremaltwomode}). To avoid these unphysical solutions we have applied the procedure described in Sec. II and derived modified extremal equations analogous to equation (\ref{EFeliminated}). We were able to construct Gröbner basis for such modified system of extremal equations, which yielded polynomial equations for the optimal $s_1$ and $s_2$ ($w\neq 0$ was assumed),
\begin{equation}
s_2\left[9s_2^4 -s_2^2 (18w^2+1)+w^2+9w^4\right]=0, \qquad s_1-s_2=0.
\label{Grobners1s2}
\end{equation}
The condition $s_1=s_2$ for the potentially optimal real $s_j$ immediately implies that it is optimal to set $s_1=s_2=0$ since any nonzero values of $s_j$ would only decrease the value of the determinant $D$. We have checked this by explicitly considering the solutions corresponding to the nonzero roots of (\ref{Grobners1s2}), which read $s_1=s_2=\pm w$ and $s_1=s_2=\pm\sqrt{1+9w^2}/3$.

\section{Summary and conclusions}

In summary, we have comprehensively investigated the maximization of the probability of preparation of non-Gaussian quantum states of light by photon counting measurements on a part of a multimode Gaussian state with vanishing coherent displacements. 
We have derived extremal equations for the parameters of the optimal Gaussian state and we have shown that these equations have the form of a system of polynomial equations. This greatly facilitates their solution and  enables us to apply techniques developed   specifically 
for solving such systems of equations. Our analysis thus provides valuable tools for design of optimal protocols and schemes for the generation of  sophisticated quantum states with complex structure in Fock basis and phase space. We have illustrated our methodology on specific examples with two and three heralding modes. 
Although the general theory was formulated for single-mode target non-Gaussian states, we have shown on a specific example of a two-mode single-rail Bell state that it can be straightforwardly generalized to multimode target states. We have also provided a procedure for engineering squeezed superpositions of Fock states, which is often desirable, for example, when one wants to  generate
 cat-like states or approximate GKP states.

It should be noted that the optimization can be in principle performed also by other numerical methods, including a naive exhaustive search when the number of parameters is small. It is, nevertheless, generally desirable to have a portfolio of various methods and approaches that can be applied to solve a given problem. Our work contributes to this portfolio for the  timely and important task of maximizing the heralding probability in optical quantum state engineering based on Gaussian resource states and photon counting.
\tcr{The main conceptual advantage of the present approach is that it allows one to perform the full global optimization of the heralding probability $p_S$ with respect to the damping parameters $X_j$. For two and three heralding modes the method appears to be  computationally very efficient and the optimization of $p_S$ for given fixed  state parameters $s_j$ and $\nu_{jk}$ can be performed on scale of seconds on an ordinary desktop machine.} We have also shown that our approach can seamlessly incorporate bounds on the available squeezing which is highly relevant in practice. The method can also be extended to states with nonzero coherent displacements. However, in such case, the expression for the normalization factor $Z$ contains an additional Gaussian function of the displacements,
 whose covariance matrix is a nontrivial function of the matrix $A$. Consequently, the resulting polynomials will be more complicated and will have higher degrees  than for the states with vanishing displacements, which can make the numerics  challenging.
 
\tcr{Even when  the damping parameters $X_j$ are optimized, the  heralding probability $p_S$  will generally decrease as the total number of the heralding photons $N$ increases. Practical applications of non-Gaussian quantum states  in optical quantum computing or quantum error correction may require close-to-deterministic state preparations. The preparation probability can be further boosted by multiplexing \cite{Scott2020, Adam2024,Forbes2025}. If a single source generates the state with probability $p_S$, then $M$ sources operating simultaneously can increase this probability to $1-(1-p_S)^M$ \cite{Forbes2025}. Additionally, quantum-state breeding schemes \cite{Etesse2015,Cotte2022,Simon2024,Sychev2017,Konno2024,Takase2024} that utilize feedforward \cite{Terhal2018,Sakaguchi2023} may lead to higher preparation probabilities and even to deterministic state generation \cite{Terhal2018}.}

\begin{acknowledgments}
This work was supported by Palacký University under Project No. IGA\_PrF\_2026\_005.
\end{acknowledgments}

\section*{Data availability statement}
The data that support the findings of this study are openly available at the following URL/DOI: \\  https://doi.org/10.5281/zenodo.20156960 \cite{FiurasekZenodo2026}.
\appendix

\begin{table}[b]
\begin{ruledtabular}
\begin{tabular}{c c  l l l  l l l}
 $n_1$ & $n_2$ & $\nu$ & $s_1$ & $s_2$ & $X_1$ & $X_2$ & $p_S$ \\ \hline
3 & 3 &	$-0.0720-1.1992i$ &  $1.3908+3.5701i$	 & $0.6509+0.0276i$ &	0.1704 &	$0.4270$ &	$0.005139$ \\
3 &	3 &	$-0.0720+1.1992i$ &  $1.3908-3.5701i$	 & $0.6509 -0.0276i$ &	0.1704 &	0.4270 &	$0.005139$ \\
3 &	3 &	$1.2394$		& $-0.9462+0.6972i$ &	$-0.9462-0.6972i$  &	0.3099	& $0.3099$ &	$0.013201$ \\
3 &	3 &	$1.2394$		& $-0.9462-0.6972i$  & $-0.9462+0.6972i$ &	0.3099 &	$0.3099$ &	$0.013201$ \\ 
3 &	3 &	$-0.0720+1.1992i$ & $0.6509-0.0276i$ &	 $1.3908-3.5701i$ &	0.4270 &	$0.1704$ &	$0.005139 $ \\
3 &	3 &	$-0.0720-1.1992i $ & $0.6509+0.0276i$ &	$1.3908+3.5701i$ &	0.4270 &	$0.1704$ &	$0.005139$ \\
\hline
4 &	2 &	$-1.0027-3.3115i$	& $-0.0720-1.1992i$ &	$13.930 +33.687i$ 	& $0.3646$ &	$0.0202$ &	$0.000088$ \\
4 & 	2 &	$-1.0027+3.3115i$	& $-0.0720+1.1992i $ &	 $13.930 -33.687i$	& $0.3646$ &	$0.0202$ &	$0.000088$ \\
4 &	2 &	$0.8587 +0.9130i$ &	 $-0.0720-1.1992i$	 & $-0.9605-0.1087i$ 	& 0.3409 &	$0.2559$ &	$0.010586$ \\
4 &	2 &	$0.8587 -0.9130i$ &	$-0.0720+1.1992i$ &	$-0.9605+0.1087i$ 	& $0.3409$ &	$0.2559$ &	$0.010586$ \\
4 &	2 &	$3.2262$	 & $1.2394$  &	$-27.769$ &	$0.3529$ &	$0.0262$ &  $0.000132$ \\
4 &	2 &	$-0.7474$	& $1.2394$	 & $4.0199$ &	$0.3890$ &	$0.1552$ &	$0.006190$ \\
\hline
4 &	3 &	$-0.0895-1.096i$	 & 0.4775-0.0499i &	$1.5633+4.4855i$ &	$0.5026$	 & $ 0.1433$ &	$0.003767$ \\
4 &	3 &	$-0.0895+1.096i$ &	0.4775+0.0499i & $1.5633-4.4855i$ &	$0.5026$ &	$0.1433$	& $0.003767$ \\
4 &	3 & $-3.1118$ & $5.6179$	& $3.3716$	& $0.1072$	& $0.1429$ &	$0.000014$ \\
4 &	3 &	$1.4362 +0.2629i$ &	$-1.0744-0.6222i$	&$ -1.4357+0.1930i$	& $0.3384$	& $0.2810$ &	$0.006458 $\\
4 &	3 &	$1.4362 -0.2629i$ &	$-1.0744+0.6222i$ &	$-1.4357-0.1930i$	& $0.3384$	& $0.2810$ &	$0.006458 $\\
4 &	3 &	$2.2701$	 & $-2.5726$ 	& $-1.7752$	& $0.1975$	& $0.2209$	& $0.000495$ \\
\hline
5 &	2 &	$-0.6915-2.6818i$ & 	$0.0286-0.8818i$	& $13.1101 +35.6363i$	& $0.4443$	& $0.0197$	& $0.000099$ \\
5 &	2 &	$-0.6915+2.6818i$	& $0.0286 +0.8818i$ & $13.1101 -35.6363i$  & $0.4443$	& $0.0197$ &  	$0.000099$ \\
5 &	2 &	$0.7487 -0.9182i$	& $0.0286 +0.8818i$	& $-1.2927+0.3643i$ & 	$0.4159$ &	$0.2202$	& $0.008903$ \\
5 &	2 &	$0.7487 +0.9182i$	& $0.0286 -0.8818i$	& $-1.2927-0.3643i$	&  $0.4159$ &	$0.2202$	& $0.008903$ \\
5 &	2 &	$2.4797$	 & $0.8686$	& $-27.003$	& $0.4400$ &	$0.0274$ &	$0.000182$ \\
5 &	2 &	$-0.7425$  &	$0.8686$	& $5.2193 $	 & $0.4762$	& $0.1244$ &	$0.005157$  
\end{tabular}
\end{ruledtabular}
\caption{In this table we specify the parameters of the $24$ configurations for generation of balanced superpositions of Fock states $\frac{1}{2}(|0\rangle+|2\rangle+|4\rangle+|6\rangle)$ and
$\frac{1}{2}(|1\rangle+|3\rangle+|5\rangle+|7\rangle)$. } 
\end{table}

\section{Coefficients of polynomials $Q_1$ and $Q_2$}
Here we provide explicit expressions for the coefficients $q_{jk}^{(1)}$ of polynomial $Q_1$ defined in equation (\ref{Qextremaltwomode}),
\begin{eqnarray}
q_{00}^{(1)}&=&2 n_1, \nonumber \\
q_{10}^{(1)}&=&-2 - 4 n_1,\nonumber   \\
q_{20}^{(1)}&=& 2 (1 + n_1) (1 - |s_1|^2 ),  \nonumber  \\
q_{01}^{(1)}&=& -4 n_1 ,  \nonumber  \\
q_{11}^{(1)}&=& 2 (1 - |\nu|^2) (1 + 2 n_1) ,   \label{q1jk}\\
q_{21}^{(1)}&=& 4 (1 + n_1) |\nu - s_1|^2,  \nonumber  \\
q_{02}^{(1)}&=& 2 n_1 (1 - |s_2|^2), \nonumber  \\
q_{12}^{(1)}&=& 2 (1 + 2 n_1)|\nu - s_2|^2 ,\nonumber  \\
q_{22}^{(1)}&=& -2 (1 + n_1)\left [4 |\nu|^2  + |s_1|^2 +|s_2|^2-|\nu|^4 - |s_1|^2|s_2|^2+ 2\Re\left(\nu^2 s_1^\ast s_2^\ast + s_1 s_2^\ast - 2 \nu (s_1^\ast + s_2^\ast)\right)\right ]
. \nonumber 
\end{eqnarray}
The expressions for the coefficients $q_{jk}^{(2)}$ of polynomial $Q_2$ can be obtained from equation  (\ref{q1jk}) by the following substitutions: $n_1 \leftrightarrow n_2$, $ s_1 \leftrightarrow s_2$, and
$jk \leftrightarrow kj$.

\section{Parameters of the configurations }

In Table I we list the parameters $\nu$, $s_1$, and $s_2$  that specify the 24 configurations that can generate the states (\ref{psi67specific}) for the four considered heralding patterns $(n_1,n_2)$. We also provide in this table the optimal values of $X_1$ and $X_2$ and the resulting maximum achievable heralding probability $p_S$ for each configuration.

\end{document}